\documentclass{jpp}
\usepackage{graphicx}
\usepackage[dvipsname]{xcolor}
\usepackage[colorlinks=true,linkcolor=olive,citecolor=olive]{hyperref}
\usepackage[utf8]{inputenc}
\usepackage[T1]{fontenc}
\usepackage{amsmath}
\usepackage{ifthen}
\usepackage{pgfplots}
\usepackage{subcaption}

\renewcommand{\vec}[1]{\mathbf{#1}}

\newcommand{\defeq}{\overset{\underset{\mathrm{def}}{}}{=}}
\newcommand{\kx}{k^x}

\newcommand{\kz}{k^z}
\newcommand{\kix}{k_i^x}
\newcommand{\kiy}{k_i^y}
\newcommand{\kiz}{k_i^z}
\newcommand{\ktx}{k_t^x}

\newcommand{\ktz}{k_t^z}
\newcommand{\vgtx}{v_{gt}^x}

\newcommand{\vgtz}{v_{gt}^z}

\newcommand{\ntx}{n_t^x}

\newcommand{\ntz}{n_t^z}
\newcommand{\ktza}{k_{t\alpha}^z}

\shorttitle{Fresnel drag in a moving magnetized plasma}
\shortauthor{J. Langlois, A. Braud and R. Gueroult}

\title{Fresnel drag in a moving magnetized plasma}

\author{J. Langlois\aff{1}
  \corresp{\email{langlois@laplace.univ-tlse.fr}},
  A. Braud\aff{1}
 \and R. Gueroult\aff{1}}

\affiliation{\aff{1}LAPLACE, Universit\'{e} de Toulouse, CNRS, INPT, UPS, 31062 Toulouse, France}

\begin{document}

\maketitle

\begin{abstract}
The change in direction of the wavevector and group velocity experienced by a wave refracted at the interface of an anisotropic medium in uniform linear motion are determined analytically. These transmission conditions, which are shown to be consistent with generalized Snell's law written in the laboratory frame, are then used to examine the effect of motion on waves incident on a magnetized plasma. For an incident wave in the plane perpendicular to the magnetic field the motion is observed to lead to non negligible deviation of the low-frequency X-mode, as well as to non-symmetrical total reflection angles. These effects are shown to be further complicated when the magnetic field is in the plane formed by the incident wavevector and the medium's velocity, as the anisotropy now competes with the motion-induced drag. Although obtained in simplified configurations, these results suggest that accounting for motion when modeling plasma waves trajectory could be important under certain conditions, calling for a more detailed quantification of the effect of motion in actual diagnostics and plasma control schemes.
\end{abstract}

\section{Introduction} \label{sec:intro}

Wave propagation in moving media is affected by motion. Sorting out these effects, one may arrange the manifestations of motion in two groups. A first group gathers manifestations which manifest as “phase” effects. One classical example is the modification of the phase index of a wave due to a motion of the medium along the wavevector, which materializes as the longitudinal Fresnel drag first postulated by~\cite{Fresnel1818} and later demonstrated by~\cite{Fizeau1851}. Another example is the phase index difference between circularly polarized modes with opposite handedness introduced by a rotational motion, from which arises polarization drag~\citep{Fermi1923,Jones1976,Player1976}. The second group gathers manifestations which manifest as “ray” effects. An example is the modification of the group velocity that leads to the transverse drag or beam deflection experienced by a wave normally incident on a moving medium~\citep{Jones1975,Player1975,Carusotto2003}.

While these effects have received considerable attention in isotropic dielectrics, notably since the realization that drag effects can be enhanced in slow light conditions~\citep{Franke-Arnold2011}, the case of magnetized plasmas has been comparatively much less studied~\citep{Gueroult2023}. On the other hand, because plasma waves are extensively used for plasma control and diagnostics, accurately modeling and quantifying the effect of motion on plasma waves is essential, and correcting for the generally neglected effect of motion might be of importance in a number of environments and applications. Information on the wave phase is for instance routinely used for diagnostics in the form of Faraday rotation measurements~\citep{Segre1999,VanEck2017}, and it has recently been surmised that rotation corrections could bear important implications for these measurements, notably in astrophysics. Specifically, interstellar magnetic field estimates are commonly inferred from pulsar polarimetry measurements that fail to account for polarization drag in the rotating magnetosphere surrounding pulsars, and neglecting this effect has recently been shown to possibly lead to errors in magnetic field estimates~\citep{Gueroult2019}. The effect of motion on the wave phase also creates opportunities to develop plasma-based non-reciprocal devices, which have been shown to hold high upside potential for light manipulation in the THz regime~\citep{Gueroult2020}. As shown above though, the effect of motion is not limited to the wave phase: it can also affect the ray dynamics. It stands to reason that the effect of motion on these rays might be equally important. For instance, given that the effectiveness of cyclotron heating~\citep{Prater2008} and current drive~\citep{Fisch1987} in tokamaks depends crucially on the accurate modeling of the propagation of radiofrequency waves across a moving plasma, it seems desirable to quantify what the effect of the plasma motion on radiofrequency beams is. Light drag is also fundamentally a manifestation of momentum coupling between the wave and the medium. Understanding how the ray dynamics in plasmas is affected by motion could thus bring insights into this basic yet important problem, notably in relation to the use of waves to drive rotation~\citep{Ochs2021,Ochs2021a,Ochs2022,Rax2023,Rax2023a,Rax2023b,Ochs2024}.

A challenge in determining the transverse drag light incident on a moving magnetized plasma experiences is that magnetized plasmas are anisotropic media. For an observer in the laboratory frame this moving anisotropic medium then appears bianisotropic~\citep{Kong1974,Kong2008}. This difference makes the analysis of refraction at a moving interface, and thus of transverse drag, more involved than in isotropic media~\citep{Player1975,Carusotto2003}. In particular, one in general has to deal with two refracted beams which are each characterized by a group velocity $\mathbf{v}_g$ that is at an angle to the propagation direction $\mathbf{k}$. To avoid this complexity, a number of studies on transverse drag in plasmas have examined particular configurations - i.e. directions of the wavevector $\mathbf{k}$, the background magnetic field $\mathbf{B}_0$ and the medium velocity $\mathbf{v}$ - for which $\mathbf{v}_g$ and $\mathbf{k}$ are aligned even if the medium is anisotropic~\citep{mukherjee1975electromagnetic,meyer1980high}. This is reminiscent of the fact that, although the dielectric tensor of a cold magnetized plasma is anisotropic, propagation parallel or perpendicular to the background magnetic field (${\mathbf{k}\parallel\mathbf{B}_0}$ or ${\mathbf{k}\perp\mathbf{B}_0}$) is characterized by a group velocity aligned with the wavevector ${\mathbf{v}_g\parallel\mathbf{k}}$~\citep[p. 145]{Ginzburg1964}. In doing so the phase index of the mode does not depend on the direction of propagation, which as we will show simplifies substantially the derivation of the beam deflection angle. Yet, one expects that the geometry in actual applications will, at least locally, not match these simplified configurations, motivating the development of a more general theory.

In this paper we address this problem by constructing a general theory for refraction at the interface with an anisotropic medium in uniform linear motion, which we use to derive a formula for the transverse drag experienced by a wave at oblique incidence on this medium. In section~\ref{sec:isotrop}, we begin by recalling the theory and manifestations of Fresnel drag at the interface with a moving isotropic media. Then, in section~\ref{sec:anisotrop}, we extend this theory to the case where the moving medium is anisotropic, and briefly show in section~\ref{sec:generalsnell} how the obtained drag coefficients can be seen as generalized Snell's laws written in the laboratory frame. We finally apply in section~\ref{sec:plasma} these new results to the case of a moving magnetized plasma, showing that drag effects could be important, especially at low frequency, and also compete with the plasma anisotropy. Section~\ref{sec:conclu} summarizes the main findings of this study.

\section{Fresnel drag in isotropic media} \label{sec:isotrop}

As a primer for the upcoming derivation in section~\ref{sec:anisotrop} of drag in moving anisotropic media, we first recall in this section the theory of drag in moving isotropic dispersive media.

A derivation of the drag experienced by an electromagnetic wave refracted at the interface of a moving isotropic dispersive media, generally referred to as Fresnel drag, was first proposed by \cite{Player1975}, and soon after that discussed in the particular case of a plasma by \cite{ko1978passage}. The central point of this derivation is, as we will show, to apply Snell's law of refraction in the frame of reference in which the interface between the medium and vacuum is at rest, using relativistic kinematics arguments to rewrite variables observed in the \textit{lab-frame} in terms of these same variables written in rest-frame of the interface. Note that because we restrict ourselves in this study to a medium in rigid body motion, the rest-frame of the interface is in fact the medium \textit{rest-frame}. We also consider here for simplicity a wave incident on a moving medium from vacuum, but note that the generalization to the case of an incident wave propagating in a material medium at rest in the lab-frame is straightforward. Lastly, when used with vector quantities the adjectives normal, longitudinal and parallel (and tangential, transverse and perpendicular) are used throughout this manuscript to indicate projections along (and in the plane perpendicular to) respectively the interface normal, the wavevector and the magnetic field.

\subsection{Relations at the interface for a moving isotropic media}

Consider a medium in uniform linear motion with velocity $\mathbf v=v\hat{\mathbf x}$ as seen in the inertial reference frame of the laboratory $\Sigma$. In its rest-frame $\Sigma'$ - which is also inertial since $\mathbf v$ is uniform - the medium is assumed to be homogeneous, isotropic and time dispersive. We note $\bar n(\omega)$ its optical index. The lab-frame coordinates system is chosen so that the interface between this moving medium (half-space ${z>0}$) and the vacuum (half-space $z<0$) defines the $(O,\hat{\mathbf x},\hat{\mathbf y})$ plane, that is that the medium moves parallel to its boundaries.

\begin{figure}
    \centering
    \includegraphics{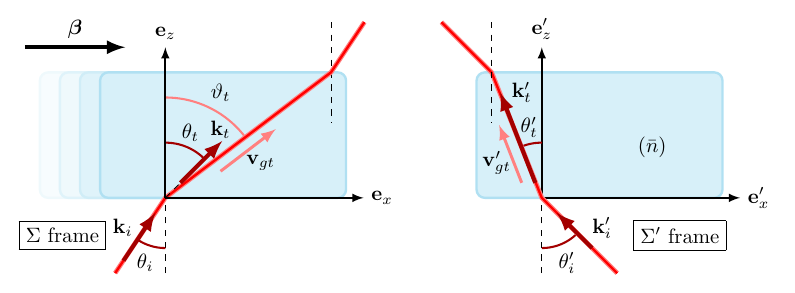}
    \caption{Light drag experienced by a wave at oblique incidence, as observed in the lab-frame $\Sigma$ (left) and in the medium's rest-frame $\Sigma'$ (right). Here the moving medium is isotropic at rest, so that the group velocity and the wavevector are aligned in $\Sigma'$. On the other hand, because they do not follow the same transformation rules from one frame to the other, they are misaligned in $\Sigma$.}
    \label{fig:problem_definition}
\end{figure}

Consider now as illustrated on the left side of figure~\ref{fig:problem_definition} an electromagnetic, monochromatic, wave with angular frequency $\omega$ in $\Sigma$ incident from the vacuum onto the moving medium. We take for simplicity this wave to be in the $(O,\hat{\mathbf x},\hat{\mathbf z})$ plane and accordingly write its wavevector $\mathbf k_i=({\kix},0,{\kiz})$, though a generalization to a non-zero $\kiy$ poses no particular problem. The angle of incidence relative to the surface normal $\hat{\mathbf z}$ in $\Sigma$ is written $\theta_i$. We would like to determine the properties of the refracted wave as seen in the lab-frame $\Sigma$. To do so we will first consider refraction as seen in $\Sigma'$.

To obtain the incident wave properties in $\Sigma'$ we use the fact that $(\omega/c,\mathbf k)$ forms the 4-wavevector in Minkowski coordinates. Its transformation to $\Sigma'$ is thus given by the Lorentz transformation \citep{einstein1905elektrodynamik,Landau1975}.
\begin{subequations} \label{eq:lorentz}
\begin{align}
    \omega' &= \gamma(\omega-{\kx} v), \\
    \Lambda: \quad {\kx}' &= \gamma\left({\kx}-{v\omega}/{c^2}\right), \\
    {\kz}' &= {\kz},
\end{align}
\end{subequations}
where $\gamma=[1-\beta^2]^{-1/2}$ is the Lorentz factor with $\beta=v/c$. In vacuum $kc=\omega$ and one immediately finds
\begin{subequations} \label{eq:incident_wave_vector_rest}
\begin{align}  
    \omega' &= \gamma \omega(1-\beta \sin \theta_i), \label{eq:incident_wave_vector_rest_omega}\\
    {\kix}' &= {\gamma\omega} \left(\sin \theta_i-\beta\right)/{c}, \\
    {\kiz}' &= {\omega} \cos \theta_i/{c}.
\end{align}
\end{subequations}

Since the medium's boundaries are at rest in $\Sigma'$, the spatial and temporal phase matching conditions of Snell's law ($\mathcal S'$) must apply in $\Sigma'$ at the interface ($z=0$) \citep{Player1975}. Using \eqref{eq:incident_wave_vector_rest}, one finds
\begin{subequations} \label{eq:snell}
\begin{align}  
    \omega_t' &= \omega', \\
    {\ktx}' & = {\kix}' = \gamma\omega\left(\sin \theta_i-\beta\right)/c.\label{eq:snell_field}
\end{align}
\end{subequations}
Here and throughout this study we denote respectively with indexes $r$ and $t$ reflected and transmitted wave variables. Equation~\eqref{eq:snell_field} provides the tangential part of the rest-frame transmitted wavevector. Meanwhile, the normal part may be determined from the definition of the optical index, that is
\begin{equation} \label{eq:rest_frame_phase_index}
    \bar n(\omega') \defeq \frac{c}{\omega'}\sqrt{{\ktx}'^2+{\ktz}'^2}.
\end{equation}

With the refracted wavevector $\mathbf{k}_t'$ in hand we can now simply use the inverse Lorentz transformation ($\Lambda^{-1}$)
\begin{subequations} \label{eq:lorentz_inverse}
\begin{align}
    \omega &= \gamma(\omega'+{\kx}'v), \\
    \Lambda^{-1}: \quad {\kx} &= \gamma({\kx}'+{v\omega'}/{c^2}), \\
    {\kz} &= {\kz}',
\end{align}
\end{subequations}
to obtain from \eqref{eq:snell}-\eqref{eq:rest_frame_phase_index} the lab-frame transmitted 4-wavevector
\begin{subequations} \label{eq:transmitted_four_vector_lab_frame}
\begin{align}
    \omega_t &= \omega, \\
    {\ktx} &= {\kix} = \omega \sin\theta_i/c, \label{eq:transmitted_four_vector_lab_frame_kx}\\
    {\ktz} &= {\ktz}' = (\gamma{\omega}/{c})\sqrt{\bar n^2(1-\beta\sin\theta_i)^2-(\sin\theta_i-\beta)^2}.
\end{align}
\end{subequations}
On the other hand, determining the refracted wave group velocity in $\Sigma$ requires a little more work. One can however take advantage here that the phase and group velocity must be collinear in $\Sigma'$, as the medium is isotropic. Specifically, one can write $\mathbf v_{g t}'=(c/\bar n_g)\hat{\mathbf k}'_t$ with $\bar n_g = \bar n + \omega'd\bar n/d\omega'$ the medium's rest-frame group-index and $\hat{\mathbf k}'_t$ the unit vector along $\mathbf k_t'$. Then, since the group-velocity transforms between inertial frames like particle velocities, the Einstein's velocity addition theorem~\citep{einstein1905elektrodynamik} can be used to give
\begin{subequations} \label{eq:transmitted_group_velocity_lab_frame}
\begin{equation}
    {\vgtx} = \frac{v_{gt}^{x\prime}+v}{1+\beta v_{gt}^{x\prime}/c } = c \frac{ {\ktx}' + \bar n_g\sqrt{{{\ktx}'}^2+{{\ktz}'}^2}}{\beta {\ktx}' + \bar n_g\sqrt{{{\ktx}'}^2+{{\ktz}'}^2}}
\end{equation}
and
\begin{equation}
    {\vgtz} = \frac 1\gamma \frac{v_{gt}^{z\prime}}{1+\beta v_{gt}^{x\prime}/c } = \frac{c}{\gamma} \frac{{\ktz}'}{\beta {\ktx}' + \bar n_g\sqrt{{{\ktx}'}^2+{{\ktz}'}^2}}.
\end{equation}
\end{subequations}
It is worth noting that using the same approach to derive the phase velocity in $\Sigma$ from its expression in $\Sigma'$ would lead to a different (an erroneous) expression compared to what is given by \eqref{eq:transmitted_four_vector_lab_frame} and the definition $\mathbf v_{\phi t} = \omega/\mathbf{k}_t$. This is simply because the phase velocity is not the spatial part of a 4-vector~\citep{DeckLeger2021}. 

Putting these pieces together, we can finally derive the angle of refraction of the wavevector $\theta_t$ and of the ray $\vartheta_t$ shown in figure~\ref{fig:problem_definition}, which from \eqref{eq:snell}-\eqref{eq:rest_frame_phase_index} and \eqref{eq:transmitted_four_vector_lab_frame}-\eqref{eq:transmitted_group_velocity_lab_frame} respectively write
\begin{equation} \label{eq:phase_angle}
    \tan \theta_t \defeq \frac{{\ktx}}{{\ktz}} = \frac{ \sin\theta_i}{\gamma\sqrt{\bar n^2(1-\beta  \sin\theta_i)^2-( \sin\theta_i-\beta)^2}}
\end{equation}
and
\begin{equation} \label{eq:group_angle}
    \tan \vartheta_t \defeq \frac{{\vgtx}}{{\vgtz}} = \gamma \frac{\bar n \bar n_{g}\beta(1-\beta \sin\theta_i)+( \sin \theta_i-\beta)}{\sqrt{\bar n^2(1-\beta \sin\theta_i)^2-( \sin\theta_i-\beta)^2}}
\end{equation}
where both $\bar n$ and $\bar n_g$ are evaluated in $\omega' = \gamma \omega(1-\beta \sin \theta_i)$. We verify that \eqref{eq:phase_angle} and \eqref{eq:group_angle} are consistent with the results derived by \cite{ko1978passage}. One further recovers the results of \cite{Player1975} in the particular case of normal incidence $\theta_i=0$. Finally these results are consistent with those of \cite{gjurchinovski2004aberration} in the limit of a nondispersive medium, i.e. for $\bar n_{g}=\bar n$.

An analogous, albeit simpler, treatment of the reflected wave shows that $\theta_r=\theta_i$, i.e. the reflection angle must always be equal to the incidence angle. 

\subsection{Refraction diagram}

Having derived expressions for the refraction angle \eqref{eq:phase_angle} and for the beam deviation angle \eqref{eq:group_angle}, we now would like to provide some physical insights into these manifestations. To do so we consider here the particular and simpler case of a non-dispersive medium. The added contribution of dispersion will be examined when discussing plasmas in section~\ref{sec:plasma}.

\begin{figure}
    \centering
    \includegraphics[width=\linewidth]{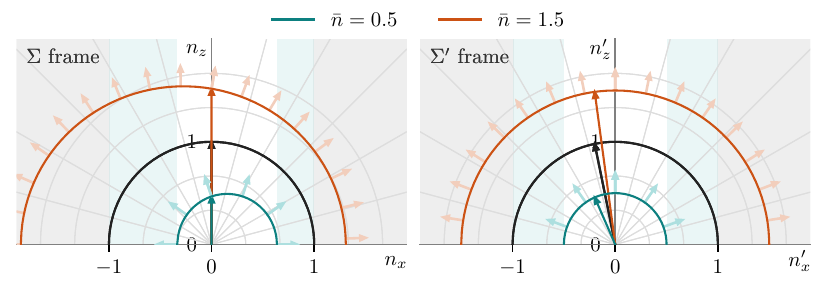}
    \caption{Refraction diagrams for a moving isotropic non-dispersive medium ($\beta=0.2$) for two values of the rest-frame refractive index $\bar{n}$. The left diagram corresponds to the lab-frame $\Sigma$, whereas the right diagram corresponds to the rest-frame $\Sigma'$. The solid black line represents the vacuum dispersion curve for the incident wave. The solid colored lines represent the moving medium dispersion curves ($\bar n>1$ in red and ${\bar n<1}$ in green). The long black and green/red arrows illustrate the incident and refracted wavevectors, respectively, for the particular case of normal incidence. The short green and red arrows on the dispersion curve represent the direction of the group velocity for the local wavevector.}
    \label{fig:refraction_diagrams_iso}
\end{figure}

To discuss the origin of drag manifestations, we use the refraction diagrams shown in figure~\ref{fig:refraction_diagrams_iso}. These diagrams, in plotting the dispersion relation in ($n_t^x,n_t^z$) space, are similar to the isofrequency diagrams used for instance by \cite{DeckLeger2021}, though in our case these diagrams further incorporate the physics of refraction at the interface. To each incident vector corresponds a point of on the unit half-circle that represents propagation in vacuum. From there conservation of the tangential wavevector component $k_i^x$ at the interface can be used to immediately deduce the wavevector component along the surface normal $k_t^z$. The refracted wave group velocity is then aligned with the normal to the dispersion curve at this point, whereas the phase velocity is aligned with the orthoradial direction. To underline the effect of motion, we consider in figure~\ref{fig:refraction_diagrams_iso} these diagrams both in $\Sigma$ and $\Sigma'$.

In $\Sigma'$ our moving medium is isotropic, so that the dispersion curve in (${n_t^x}',{n_t^z}'$) is also a half-circle, with radius $\bar{n}$. Equivalently the phase and group velocity are aligned, as expected for an isotropic medium. This is illustrated in the refraction diagrams on the right hand side in figure~\ref{fig:refraction_diagrams_iso}. In contrast, we see moving to the  refraction diagrams on the left hand side in figure~\ref{fig:refraction_diagrams_iso} that the dispersion curve in ($n_t^x,n_t^z$) is distorted. This is the effect of motion.  One finds in particular that the dispersion curve is no longer a half-circle, and is notably asymmetrical with respect to $k_i^x$. As a result there is now a deviation between the phase and group velocity. This is light dragging induced by motion.

Trying to shed light onto the origins of these manifestations, a contribution to the distortion of the dispersion curve going from $\Sigma'$ to  $\Sigma$ is the misalignment between the wavevectors expressed in these two frames. The latter is itself a result of relativistic aberration, as can be seen in the Lorentz transformation \eqref{eq:lorentz}. Specifically, since we find Snell's laws to take their usual form in the rest-frame $\Sigma'$, the incident wavevector for which a symmetrical dispersion curve is found is $\vec k_i'$, which is not aligned with $\vec k_i$ as the relative motion introduces a change of the component along the motion direction. This misalignment carries back to $\Sigma'$ through the inverse Lorentz transformation written for $\vec k_t'$, giving the dispersion curve in ($n_t^x,n_t^z$) space an asymmetrical shape. 

Finally, another facet of this motion induced asymmetry is the shift of the critical angles above which total reflection occurs, as can be seen on the left-hand-side in figure~\ref{fig:refraction_diagrams_iso} for $\bar{n}<1$. For an isotropic medium at rest, the total reflection critical angles are symmetrical $\pm{\theta_i^c}'$. This property comes naturally on the right-hand-side in figure~\ref{fig:refraction_diagrams_iso} as the dispersion curve in $\Sigma'$ is again a half-circle centered on the origin. However, we verify on the left-hand-side in figure~\ref{fig:refraction_diagrams_iso} that the shift of the dispersion curve due to the motion 
leads to total reflection critical angles that are no longer symmetrical. 

\section{Refraction laws for an anisotropic dispersive media in uniform linear motion} \label{sec:anisotrop}

Having reviewed in the previous section the main traits of Fresnel drag in a moving isotropic medium, we can now examine how these results generalize to the case of an anisotropic medium. The major modification introduced by the medium's \textit{material} anisotropy lies in the loss of the collinearity between phase and group velocities in the rest-frame $\Sigma'$, which as we have shown above was an essential argument in determining interface relations for the moving isotropic medium. An added complexity is that an anisotropic medium supports several propagation modes (denoted by an index $\alpha$ here), so that a monochromatic incident ray is refracted into as many rays as there are modes.

Our starting point is the \textit{dispersion function} for the mode $\alpha$ in the medium at rest, which we write here in generality
\begin{equation} \label{eq:dispersion_function}
    \mathcal D'_\alpha\left(\omega',\mathbf k'\right)=\bar n_\alpha\left(\omega',\hat{\mathbf k}'\right)-k'c/\omega'.
\end{equation}
We underline here that, importantly and unlike the isotropic case, $\bar n_\alpha$ is now a function of the direction of the wavevector $\hat{\mathbf k}'$. The dispersion function of the different modes then yield the \textit{dispersion relation}
\begin{equation}
\label{Eq:dispersion_rel}
    \prod_\alpha \mathcal D'_\alpha\left(\omega',\mathbf k'\right)=0.
\end{equation}
The dispersion relation \eqref{Eq:dispersion_rel} is classically used to determine $\omega'$ for a given $\mathbf k'$, or vice-versa. Here, however, we have seen in \eqref{eq:snell} that the continuity conditions at the interface impose $\omega'=\gamma \omega(1-\beta  \sin \theta_i)$ and ${\ktx}'=\gamma\omega(\sin \theta_i-\beta)/c$. The dispersion relation \eqref{Eq:dispersion_rel} is then instead used to recast \eqref{eq:dispersion_function} {as an implicit equation $\mathcal D'_\alpha\left(\omega',{\ktx}',k_{t\alpha}^{z\prime}\right)=0$ for the normal component of the refracted wavevector $k_{t\alpha}^{z\prime}$ as a function of the tangential component of the refracted wavevector ${\ktx}'$ and the rest-frame frequency $\omega'$, which we write
\begin{equation} \label{eq:k_def}
    k_{t\alpha}^{z\prime} \defeq \mathcal K_\alpha'(\omega',{\ktx}').
\end{equation}
In other words
\begin{equation} \label{eq:k_def}
    D'_\alpha\left(\omega',{\ktx}',\mathcal K_\alpha'(\omega',{\ktx}')\right)=0.
\end{equation}
}
One further verifies that the inverse Lorentz transformation \eqref{eq:lorentz_inverse} does not affect this wavevector component, that is to say $k_{t\alpha}^z=k_{t\alpha}^{z\prime}$, which gives 
\begin{equation} \label{eq:k_def_lab}
    {\ktza} = \mathcal K_\alpha'(\omega',{\ktx}').
\end{equation}

Equipped with the tangential and normal lab-frame wavevector components \eqref{eq:transmitted_four_vector_lab_frame_kx} and \eqref{eq:k_def_lab}, respectively, and the Doppler shifted frequency \eqref{eq:incident_wave_vector_rest_omega}, the lab-frame transmitted phase angle \eqref{eq:phase_angle} writes
\begin{equation} \label{eq:theta_alpha}
    \tan \theta_{t\alpha} = \frac{\omega \sin\theta_i}{c \mathcal K'_\alpha\left(\omega', {\ktx}'\right)}.
\end{equation}
We stress that $\omega'$ and ${\ktx}'$ in \eqref{eq:theta_alpha} are not free variables but instead known functions of the lab-frame incident angle $\theta_i$ and the lab-frame wave frequency $\omega$ through \eqref{eq:transmitted_four_vector_lab_frame_kx} and \eqref{eq:incident_wave_vector_rest_omega}. 

Considering now the group velocity of the refracted wave, we can no longer take advantage of the alignment of the phase and group velocities in the rest-frame $\Sigma'$ used above for isotropic media. Instead we use here the definition \eqref{eq:group_angle} together with the inverse Lorentz transformation \eqref{eq:lorentz_inverse} to formally express the beam deviation as a function of the rest-frame wavevector and wave frequency, giving
\begin{align} 
    \tan \vartheta_t &= - \frac{\partial {\ktz}}{\partial {\ktx}} \nonumber\\
    &= -\left[\frac{\partial {\ktz}'}{\partial {\ktx}'}\frac{\partial {\ktx}'}{\partial {\ktx}}+\frac{\partial {\ktz}'}{\partial \omega'}\frac{\partial \omega'}{\partial {\ktx}} \right ] \nonumber\\
    &= \gamma\left[v \frac{\partial {\ktz}'}{\partial \omega'} - \frac{\partial {\ktz}'}{\partial {\ktx}'}\right ].
\end{align}
Plugging in \eqref{eq:k_def} then yields
\begin{equation} \label{eq:vartheta_alpha}
    \tan \vartheta_{t\alpha} = \gamma\left[v \frac{\partial \mathcal K_\alpha'}{\partial \omega'}(\omega',{\ktx}') - \frac{\partial \mathcal K_\alpha'}{\partial {\ktx}'}(\omega',{\ktx}')\right ]
\end{equation}
where again $\omega'$ and ${\ktx}'$ are given by \eqref{eq:transmitted_four_vector_lab_frame_kx} and \eqref{eq:incident_wave_vector_rest_omega}.

In magnetized plasmas the dispersion relations are often expressed in terms of the components of the refractive index parallel and perpendicular to the background magnetic field $n_\parallel$ and $n_{\perp}$. Pursuing this analogy, we can similarly recast our results in terms of the refractive indexes parallel and perpendicular to the motion ${\ntx}'=c{\ktx}'/\omega'$ and ${\ntz}'=\mathcal N_\alpha' = c\mathcal K_\alpha'/\omega'$, which leads to
\begin{equation} \label{eq:transmited_phase_angle_lab}
    \tan \theta_{t\alpha} = \frac{ \sin\theta_i}{\mathcal N'_\alpha\left(\omega', {\ntx}'\right)}
\end{equation}
and
\begin{equation} \label{eq:transmited_group_angle_lab}
    \tan \vartheta_\alpha = \gamma\left[\beta \left(\mathcal N_\alpha'(\omega',{\ntx}')+\omega'\frac{\partial \mathcal N_\alpha'}{\partial \omega'}(\omega',{\ntx}')\right) - \frac{\partial \mathcal N_\alpha'}{\partial {\ntx}'}(\omega',{\ntx}')\right ]
\end{equation}
where now
\begin{equation} \label{eq:ntx_prime}
    {\ntx}' = \frac{\sin \theta_i-\beta}{1-\beta\sin \theta_i}.
\end{equation}

We note that in the particular case where the rest-frame refractive index $\bar n_\alpha$ does not depend of the propagation direction, as it is notably the case for O-X and L-R modes in magnetized plasmas~\citep{Ginzburg1964}, \eqref{eq:k_def} reduces to
\begin{equation}
    \mathcal K_\alpha'(\omega',{\ktx}') = \sqrt{\left(\frac{\omega' \bar n_\alpha(\omega')}{c}\right)^2-{{\ktx}'}^2 }.
\end{equation}
One verifies that plugging this results into \eqref{eq:theta_alpha} and \eqref{eq:vartheta_alpha} yields the lab-frame transmitted angles \eqref{eq:phase_angle} and \eqref{eq:group_angle} of an isotropic medium with refractive index $\bar n_\alpha$. Drag phenomena for this particular mode thus manifest essentially as in an isotropic medium with the appropriate index.

\section{Generalized Snell's laws in lab-frame} \label{sec:generalsnell}

\begin{figure}
    \centering
    \includegraphics{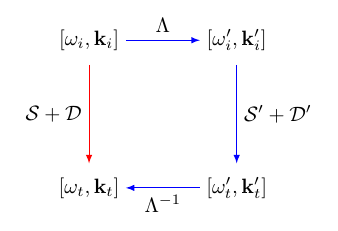}
    \caption{Representation of the two possible paths to derive the 4-wavevector of a beam refracted at the interface with a moving medium. The blue path represent the approach originally proposed by \cite{Player1975}, which uses standard refraction laws written in the rest-frame (see sections~\ref{sec:isotrop} and \ref{sec:anisotrop}). The red path directly employs generalized Snell's laws written in the lab-frame, in which motion appears as an effective property.}
    \label{fig:paths}
\end{figure}

The derivation of drag phenomena proposed in sections ~\ref{sec:isotrop} and \ref{sec:anisotrop} followed the approach proposed \cite{Player1975}, that is to say to derive lab-frame refraction properties from those determined in the medium rest-frame. This is the path illustrated in blue in figure~\ref{fig:paths}. A different approach consists in modeling the interface directly in the lab-frame. Indeed, since the moving medium as seen from the lab-frame appears to have been bestowed additional properties as a result of motion (for instance bianisotropy~\citep{Kong2008} and spatial dispersion~\citep{lopez1997dispersion}), one can seek to recast these additional properties in the form of generalized reflection and refraction laws, which can then be directly applied in the lab-frame~ \citep{pyati1967reflection,kong1968wave,mukherjee1975electromagnetic,huang1994reflection}. This is the red path in figure~\ref{fig:paths}. In this section we make a brief digression to present this second approach and underline the equivalence of the two approaches.

As we have seen in section~\ref{sec:isotrop}, the relation for the tangential part of the refracted wavevector is entirely deduced from continuity at the interface. On the other hand, determining its normal component demands the dispersion function for the wave in the medium. Generalizing this to the lab-frame thus entails determining an effective dispersion function $\mathcal D(\omega,\vec k)$, that is the dispersion function that characterizes wave propagation as seen in the lab-frame. This is simply a generalization of the previously defined rest-frame dispersion function \eqref{eq:dispersion_function}. Here we use the result derived by \cite{censor1980dispersion} that the dispersion relations of any linear medium are covariant between inertial frames. This means that $\mathcal D(\omega,\vec k)$ can simply be obtained by substituting to $\vec k'$ and $\omega'$ in $\mathcal D'(\omega',\vec k')$ their Lorentz transformation in the lab-frame, or mathematically
\begin{equation}\label{eq:D=D'}
    \mathcal D(\omega,\vec k)=\mathcal D'\big(\omega'[\omega,\vec k],\vec k'[\omega,\vec k]\big).
\end{equation}
Given this lab-frame dispersion function, Snell's law for the normal component of the wavevector then writes 
\begin{equation}
{{\ktz}=\mathcal K(\omega_t,{\ktx})}
\end{equation}
where $\mathcal K$ is similarly the analog of $\mathcal K'$ in \eqref{eq:k_def}. The generalized Snell's laws ($\mathcal S$) in the lab-frame thus write
\begin{subequations}
    \begin{align}
        \omega_t&=\omega,\\
        {\ktx}&={\kix},\\
        {\ktz}&=\mathcal K(\omega_t,{\ktx})\label{eq:generalized_snell_normal}.
    \end{align}
\end{subequations}

Since the equivalence of the two methods has already been demonstrated for the tangential part of the wavevector, all that is left to do is to verify the consistency of the two methods for the normal component of the wavevector, that is \eqref{eq:generalized_snell_normal}. Going back to our rest-frame analysis, the rest-frame refracted wavevector $\vec k_t'$ verifies the rest-frame dispersion relation, that is   
\begin{equation}
\label{Eq:dispersion_snell_rest_frame}
    \mathcal D'\left(\omega'_t,{\ktx}',{\ktz}'=\mathcal K'(\omega'_t,{\ktx}')\right)=0.
\end{equation}
Meanwhile, substituting the lab-frame Lorentz transformation variables, and noting importantly that the normal component of the wavevector is unaffected ${\ktz}={\ktz}'$, one can write
\begin{align}
    \mathcal D'\left(\omega'_t,{\ktx}',{\ktz}'=\mathcal K'(\omega'_t,{\ktx}')\right) 
    &=\mathcal D'\left(\omega'_t[\omega_t,{\ktx}],{\ktx}'[\omega_t,{\ktx}],{\ktz}'={\ktz}\right) \nonumber\\
    &=\mathcal D\left(\omega_t,{\ktx},{\ktz}\right).\label{Eq:Censor_eq}
\end{align}
Putting together \eqref{Eq:dispersion_snell_rest_frame} and \eqref{Eq:Censor_eq}, this implies that ${\ktz}$ verifies the dispersion relation in the lab-frame, or in other words that ${\ktz}=\mathcal K(\omega_t,{\ktx})$, which is precisely \eqref{eq:generalized_snell_normal}. The two methods, as illustrated in blue and red in figure~\ref{fig:paths}, are thus indeed equivalent.

To summarize, for a medium with known rest-frame dispersion function $\mathcal D'(\omega',\vec k')$, the covariance of the dispersion function demonstrated by \cite{censor1980dispersion} can be used to obtain the refracted wavevector directly in the lab-frame. This result is consistent with the rest-frame approach proposed by~\cite{Player1975} and used above in sections~\ref{sec:isotrop} and \ref{sec:anisotrop}.

\section{Application to moving magnetized plasmas} \label{sec:plasma}

With the theory for the drag induced by a moving anisotropic media in hand, we can now examine more particularly how these effects manifest themselves in a magnetized plasma in uniform linear motion with respect to the observer. We consider first these effects for the ordinary and extraordinary modes classically obtained for propagation perpendicular to the magnetic field, as the collinearity of rest-frame phase and group velocities~\citep{Ginzburg1964} simplifies algebra. In doing so, we recover previously established results, but also underline the important contribution of dispersion and Doppler shift to dragging effects. We then examine the more general case, making full use of the results from section~\ref{sec:anisotrop}. 

\subsection{Magnetic field normal to the incidence plane (O and X modes)}

\begin{figure}
    \centering
    \includegraphics{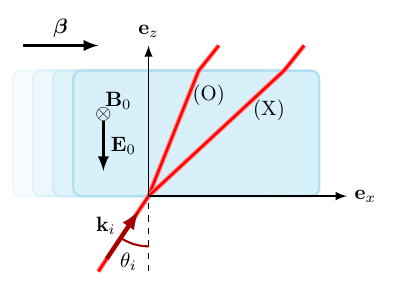}
    \caption{\label{fig:plasma_config_a} The incident wave is in the plane normal to the background magnetic field permeating the magnetized plasma in uniform linear motion. The rest-frame modes are the classical ordinary (O) and extraordinary (X) waves.}
\end{figure}

We consider here the moving medium to be a magnetized plasma with background magnetic field $\mathbf{B}_0'=B_0'\mathbf{\hat{y}}$ in its rest-frame. Given our choice to have the incident wavevector $\mathbf{k}_i$ in the $(O,\mathbf{\hat{x}},\mathbf{\hat{z}})$ plane, and the result that the Lorentz transformed $\mathbf{k}_i'$ for $\mathbf{v} = v\mathbf{\hat{x}}$  is also in the $(O,\mathbf{\hat{x}},\mathbf{\hat{z}})$ plane, this corresponds as shown in figure~\ref{fig:plasma_config_a}  to perpendicular propagation in the rest-frame $\Sigma'$. The normal modes are hence classically the ordinary ($\rm{O}$) and extraordinary ($\rm{X}$) waves~\citep{rax2005physique}. We note that given the Lorentz transformations of fields
\begin{subequations} \label{eq:field_lorentz_transform}
\begin{align} 
    \mathbf{E}_\parallel &= \mathbf{E}_\parallel', \\
    \mathbf{B}_\parallel &= \mathbf{B}_\parallel', \\
    \mathbf{E}_\bot &= \gamma \left( \mathbf{E}_\bot' - \mathbf{v} \times \mathbf{B}' \right), \\
    \mathbf{B}_\bot &= \gamma \left( \mathbf{B}_\bot' + {\mathbf{v}}/{c^2} \times \mathbf{E}' \right),
\end{align}
\end{subequations}
this field configuration implies ${\mathbf{B}_0 = \gamma \mathbf{B}_0' \sim \mathbf{B}_0'}$ and ${\mathbf{E}_0=-\mathbf{v} \times \mathbf{B}_0}$ in the lab-frame. This lab-frame field configuration ($\mathbf{E}_0,\mathbf{B}_0$) is consistent with a plasma drift with velocity $\mathbf{v}$.

\subsubsection{O-mode}

The refractive index of the O-mode in the rest-frame, whose polarization is along the magnetic field $\mathbf{B}_0'$, simply writes
\begin{equation}\label{eq:index_0}
    \bar n_{\textrm{O}}(\omega')=\sqrt{1-\left(\frac{\omega_{p}'}{\omega'}\right)^2}
\end{equation}
where $\omega_{p}'^2=\sum_s \omega_{ps}'^2=\sum_s n_s' e^2/(m_s'\epsilon_0)$ is the rest-frame plasma frequency. The associated dispersion function is
\begin{equation} \label{eq:o_mode_invariance}
    \mathcal D_\textrm{O}'(\omega',\mathbf k')=\omega'^2-k'^2c^2-\omega_{pe}'^2,
\end{equation}
which, noting that the plasma frequency is Lorentz invariant $\omega_{pe}=\omega_{pe}'$~\citep{chawla1966note},  can be rewritten using the Lorentz transformation for $\omega'$ and $\mathbf k'$ as the lab-frame dispersion function
\begin{equation} \label{eq:o_mode_invariance_2}
    \mathcal D_\textrm{O}(\omega,\mathbf k)=\omega^2-k^2c^2-\omega_{pe}^2.
\end{equation}

Comparing \eqref{eq:o_mode_invariance} and \eqref{eq:o_mode_invariance_2} shows that the O-mode dispersion relation is remarkably Lorentz-invariant. As a consequence, it is as noticed by \cite{mukherjee1975electromagnetic} and \cite{ko1978passage} unaffected by motion. This property, which is characteristic of modes satisfying $\bar n \bar n_g=1$, was suggested to support Minkowski's formulation of momentum partitioning in a medium~\citep{arnaud1976dispersion,jones1978radiation}. In this case the relations at the interface simply follow the static Snell's laws. One indeed finds plugging the mode's refractive index \eqref{eq:index_0} into \eqref{eq:transmited_phase_angle_lab}-\eqref{eq:transmited_group_angle_lab} that $\bar n_\textrm{O}\sin\theta_t=\sin\theta_i$ and $\vartheta_t=\theta_t$, that is to say that the group velocity remains aligned with the wavevector in the lab-frame. 

\subsubsection{X-mode}

The dispersion relation of the X-mode, whose polarization is in the $(\mathbf{v}, \mathbf{k'})$ plane perpendicular to $\mathbf{B}_0'$, writes~\citep{rax2005physique}
\begin{subequations} \label{eq:xmode}
\begin{equation}\label{eq:index_X}
    \bar n_\textrm{X}(\omega')=\sqrt{\frac{(\omega'^2-\omega_\textrm{L}'^2)(\omega'^2-\omega_\textrm{R}'^2)}{(\omega'^2-\omega_\textrm{UH}'^2)(\omega'^2-\omega_\textrm{LH}'^2)}}
\end{equation}
where
\begin{equation}
    \omega_\textrm{R/L}' = \mp \frac{\Omega_{ce}'+\Omega_{ci}'}{2} + \frac{1}{2}\sqrt{(\Omega_{ce}'-\Omega_{ci}')^2+4\omega_{p}'^2}
\end{equation}
are the right and left cutoffs and
\begin{equation}
    \omega_\textrm{UH/LH}' = \left[\frac{\varpi_e'^2+\varpi_i'^2}{2}\pm \frac{1}{2}\sqrt{(\varpi_e'^2-\varpi_i'^2)^2+4\omega_{pe}'^2\omega_{pi}'^2}\right]^{1/2}
\end{equation}
\end{subequations}
are the upper- and lower-hybrid frequencies, and where we write $\varpi_s'^2 = \omega_{ps}'^2+\Omega_{cs}'^2$ with $\Omega_{cs}'=q_sB_0'/m_s'$ the signed rest-frame cyclotron frequency for species $s$. One verifies that $\bar n \bar n_g\neq1$, so that the X-mode is expected to experience Fresnel drag. 

\textbf{Asymptotic trends.} Since the wave index \eqref{eq:index_X} does not depend on the incidence angle $\theta_i$, drag phenomena for the X-mode can still, as already indicated above, be evaluated with the comparatively simpler isotropic model presented in section~\ref{sec:isotrop}. Specifically plugging \eqref{eq:index_X} into \eqref{eq:phase_angle}-\eqref{eq:group_angle} yields lengthy yet analytical formulae for the refraction and group velocity angles  for any lab-frame wave frequency $\omega$. As is customary, simpler forms can however be obtained if considering separately different wave frequency bands. For instance, if one focuses on the high-frequency electronic response, the wave index reduces to
\begin{equation}
    \bar n_\textrm{X}(\omega'\gg\omega_\textrm{LH}') \sim \sqrt{1-\frac{\omega_{pe}'^2(\omega'^2-\omega_{pe}'^2)}{\omega'^2(\omega'^2-\omega_{pe}'^2-\Omega_{ce}'^2)}}.
\end{equation}
In this limit \eqref{eq:phase_angle}, combined with the trigonometric relation $\sin{\theta}=\tan{\theta}/\sqrt{1+\tan^2{\theta}}$ and $\Omega_{cs}=\Omega_{cs}'/\gamma$ here, gives 
\begin{equation}
    \sin\theta_t=\sin\theta_i\left[1-\frac{\omega_{pe}^2}{\omega^2}\frac{(1-\beta^2)\omega_{pe}^2-(1-\beta\sin\theta_i)^2\omega^2}{\Omega_{ce}^2+(1-\beta^2)\omega_{pe}^2-(1-\beta\sin\theta_i)^2\omega^2}\right]^{-1/2}
\end{equation}
which we verify is precisely the generalized Snell's law derived by~\cite{mukherjee1975electromagnetic}. Furthermore, using this same high-frequency wave index in the relation for the group velocity \eqref{eq:group_angle} in the limit of normal incidence $\theta_i=0$ leads to
\begin{equation}
\label{Eq:drag_X_high}
    \tan\vartheta_t = \frac{\beta \gamma \omega_{pe}^2\Omega_{ce}^2}{\sqrt{[\omega^2-\omega_{pe}^2/\gamma^2-\Omega_{ce}^2]^3[(1-\omega_{pe}^2/\omega^2)(\gamma^2\omega^2-\omega_{pe}^2)-\gamma^2\Omega_{ce}^2]}},
\end{equation}
which we verify is the result derived by \cite{meyer1980high}. At very high frequency $\omega\gg\omega_{pe},\Omega_{ce}$, $\tan\vartheta_t\propto\beta/(\gamma^2\omega^4)$. The drag is hence very small short of relativistic velocities. 

The low frequency regime $\omega'\leq\omega_\textrm{LH}'$ has in contrast to our knowledge received less attention. In the limit $\omega'\leq\Omega_{ci}'$ and assuming $v_\textrm{A}'/c=\Omega_{ci}'/\omega_{pi}'\ll1$ with $v_\textrm{A}'$ the Alfv{\'en} velocity one classically shows that 
\begin{equation}
    \bar n_\textrm{X}(\omega'\leq\omega_\textrm{LH}') \sim \frac{c}{v_\textrm{A}'}\sqrt{1-\left(\frac{\omega'}{\Omega_{ci}'}\right)^2}.
\end{equation}
Equation \eqref{eq:group_angle} then gives in the very low frequency $\omega'\ll\Omega_{ci}'$ and for normal incidence 
\begin{equation}
\label{Eq:drag_X_low}
    \tan\vartheta_t = \frac{v}{v_\textrm{A}}
\end{equation}
to lowest order in $\beta$. This shows that, in contrast with the high frequency regime, non-negligible drag can occur for the compressional Alfv{\'e}n (or fast magnetoacoustic) branch of the X-mode, and that even for non-relativistic velocities.

To confirm these trends, explore the intermediate frequency regimes and study the effect of incidence, we now examine the results obtained using the full solution \eqref{eq:index_X} in \eqref{eq:phase_angle}-\eqref{eq:group_angle}. To this end we consider as a baseline a hydrogen plasma with density ${n_e=10^{19}\textrm{ m}^{-3}}$ and rest-frame magnetic field ${B_0'=1\textrm{ T}}$.

\begin{figure}
    \centering
    \includegraphics[]{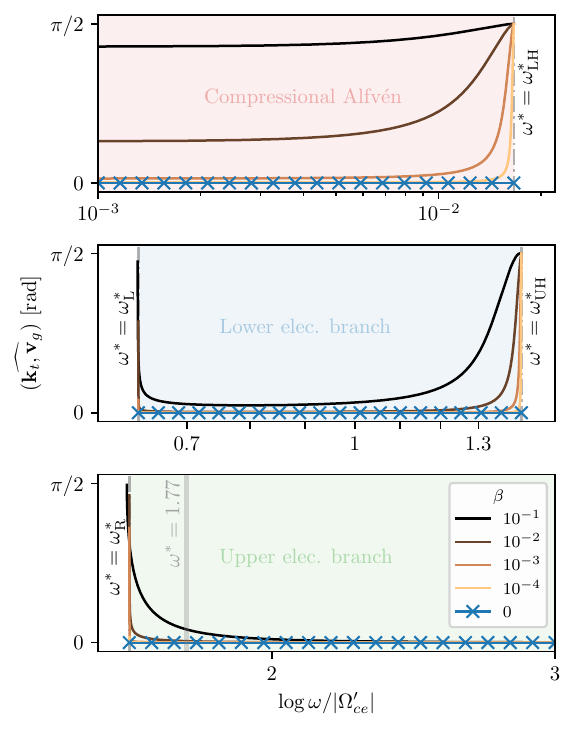}
    \caption{Angle between the group velocity $\mathbf v_g$ and the wavevector $\mathbf k_t$ of the X-mode refracted by a wave at normal incidence on a moving magnetized plasma, as a function of the frequency and for several values of the velocity, for a hydrogen plasma with $n_e=10^{19}\textrm{ m}^{-3}$ and $B_0'=1\textrm{ T}$. Here ($\mathbf v, \mathbf k_t, \mathbf B_0$) forms an orthogonal basis. The three panels represent the three standard propagation branches of the X-mode. The superscript~${}^*$ indicates a normalization by the rest-frame electron cyclotron frequency $|\Omega_{ce}'|$. The vertical gray line for $\omega^*=1.77$ on the third panel highlights the frequency for which oblique incidence is examined in figure~\ref{fig:refraction_diagram_Xmode}. }
    \label{fig:Xmode}
\end{figure}

\textbf{Normal incidence.} To start with, we consider the case of normal incidence, that is when ($\mathbf{v},\mathbf{B_0}',\mathbf{k}_i$) forms an orthogonal basis. In this particular case the transmitted wavevector $\mathbf{k}_t$ is conveniently along $\mathbf{k}_i$, i.e. along $\mathbf{\hat{z}}$. Figure~\ref{fig:Xmode} plots the angle $(\widehat{\vec k_t,\vec v_g})$ between the refracted wavevector $\vec k_t$ and the group velocity $\vec v_g$ across the entire frequency range for different values of $\beta=v/c$. Without motion, i.e. for $\beta=0$, we recover the classical behavior of the X-mode, that is to say a group velocity that is aligned with the wavevector ${\vec v_g\parallel\vec k}$. This materializes in figure~\ref{fig:Xmode} as an angle $(\widehat{\vec k_t,\vec v_g})$ that is zero for all frequencies. We also recognize in figure~\ref{fig:Xmode} for $\beta=0$ the three usual propagation branches of this mode, namely below the lower-hybrid resonance $\omega_\textrm{LH}$, in between the left cutoff $\omega_\textrm{L}$ and the upper-hybrid resonance $\omega_\textrm{UH}$, and finally above the right cutoff $\omega_\textrm{R}$~\citep{rax2005physique}.

Moving on to the effect of velocity, figure~\ref{fig:Xmode} confirms that the angle $(\widehat{\vec k_t,\vec v_g})$ is now finite for $\beta\neq0$, and positive for all frequencies. This means that the X-mode is dragged in the direction of motion. This drag is further found to increase with velocity for all frequencies. Overall, drag effects are observed to be strongly enhanced near resonances and cutoffs. They even reach at these frequencies the maximum angle $\pi/2$ which represents a limit case where the wave is fully dragged by the medium. Although this increase appears consistent with the classic $\bar{n}_g-1/\bar{n}$ scaling of transverse drag~\citep{Player1975} (the phase velocity goes to zero near cutoffs while the group velocity goes to infinity near resonances) and with the observation of enhanced drag effects in slow-light media~\citep{Franke-Arnold2011}, results for these frequencies warrant caution as the cold plasma model used in this study is expected to break down. Notwithstanding these limitations, these results suggest that augmented drag effects could be achieved near resonances and cutoffs. 

Away from resonances and cutoffs, figure~\ref{fig:Xmode} confirms that drag effects are negligible for the two high frequency bands, consistent with \eqref{Eq:drag_X_high}. On the other hand, figure~\ref{fig:Xmode} also confirms that significant drag occurs at low frequency $\omega\leq\omega_\textrm{LH}$ as anticipated from \eqref{Eq:drag_X_low}. In this low frequency regime the angle $(\widehat{\vec k_t,\vec v_g})$ is observed to be nearly independent of the wave frequency, consistent with the fact that the X-mode at these frequencies is nearly non-dispersive. Quantitatively, we find a drag of a few degrees for $\beta\sim 10^{-4}$. This result matches the prediction from \eqref{Eq:drag_X_low} as $c/v_A\sim40$ for the plasma parameters considered here, and larger drags are expected for denser plasmas at the same field (or a similar density but at weaker fields).

\textbf{Oblique incidence and Doppler.} While the low-frequency X-mode is nearly non-dispersive, this is not the case at higher frequency, which brings additional complexity. More precisely, dispersion can manifests due to the Doppler shift experienced by the wave as seen in the rest-frame, leading to new effects compared to the nondispersive medium considered in section~\ref{sec:isotrop}. Specifically, the Lorentz transformation for the wave frequency \eqref{eq:incident_wave_vector_rest_omega} shows that the rest-frame frequency $\omega'$ depends on the angle of incidence $\theta_i$. As a result, the propagation bands of the X-mode, which normally are independent of the wavevector direction, now depend on the incidence angle $\theta_i$.

\begin{figure}
    \centering
    \includegraphics[width=10.5cm]{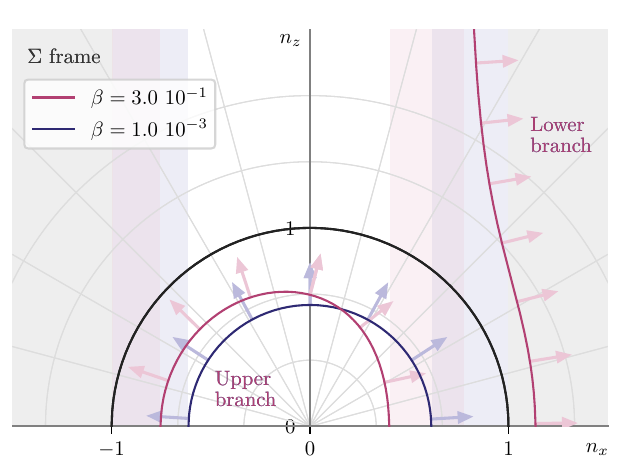}
    \caption{Refraction diagrams for a wave with rest-frame right cutoff ($\omega/|\Omega'_{ce}|=1.77$) at oblique incidence on a moving plasma for two different values of $\beta$. Absent motion the refracted wave is on the upper branch of the X-mode. The colored bands highlight the regions of total reflection.}
    \label{fig:refraction_diagram_Xmode}
\end{figure}

To illustrate this point, we have plotted in figure~\ref{fig:refraction_diagram_Xmode} the refraction diagrams obtained for the X-mode with normalized rest-frame wave frequency $\omega^*=\omega/|\Omega_{ce}'|= 1.77$. As shown in figure~\ref{fig:Xmode} this frequency falls in the high-frequency electronic branch of the X-mode for $\beta=0$, just above the right cutoff.  We see in figure~\ref{fig:refraction_diagram_Xmode} that while the effect of motion is limited at low $\beta$, new features emerge for larger $\beta$. It is notably found that the incident wave, which again verifies $\omega>\omega_\textrm{R}$, can in fact couple instead to the low frequency electronic branch for sufficiently large $\beta$ and $k_i^x$ (i.e. sufficiently large incidence angle $\theta_i$). For $k_i^x>0$ the Doppler shift is indeed such that $\omega'<\omega$, and $\omega'$ can become smaller than the rest-frame upper-hybrid frequency $\omega_\textrm{UH}'$. In this case a strong drag is observed, as $\omega'$ is close to the resonance. As a direct consequence of this behavior, we note the existence of an intermediate total reflection region in between the values of $k_i^x$ yielding these two distinct branches. This remarkable feature is entirely due to the plasma motion. Note also that a symmetrical behavior can be observed if choosing a rest-frame frequency in the low frequency electronic branch just below the upper-hybrid resonance, and this time $k_i^x<0$ so that $\omega'>\omega$.

To sum up, it is found here that while drag effects on the X-mode are generally small at high frequency away from cutoffs and resonances, they can be significant at low frequency, i.e. for the compressional Alfv{\'en} branch. In addition, the motion can have a noticeable effect near resonances and cutoffs, where it can lead to jumps from a given branch to the other as the incidence angle changes at fixed wave frequency, to the onset of incidence angle-dependent asymmetric propagation windows, and also possibly to augmented drag effects.  

\subsection{Magnetic field in the incidence plane}

\begin{figure}
    \centering
    \includegraphics[]{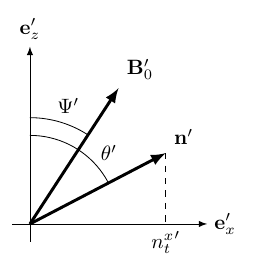}
    \caption{\label{fig:angles} Configuration with the rest-frame magnetic field $\mathbf{B}_0'$ in the plane defined by the incident wavevector $\mathbf{k}_i$ and the medium velocity $\vec v$. Propagation is in this case in general oblique. }
\end{figure}

To expose how rest-frame anisotropy can further complicate the drag picture, we finally consider the case where the magnetic field $\mathbf{B}_0'$ lies in the plane defined by the incident wavevector $\mathbf{k}_i$ and the medium velocity $\vec v$. As depicted in figure~\ref{fig:angles}, we write $\Psi'$ the angle in this plane between the normal to the velocity and the magnetic field. Other than for singular values of $\Psi'$ the wavevector is now inclined with respect to the rest-frame magnetic field, so that the rest-frame indexes $\bar{n}_{\alpha}'$ indeed depend on the wavevector of the refracted wave $\vec k_t'$, i.e. on the propagation direction in the rest-frame.

\subsubsection{General formulation}

Propagation in the rest-frame is oblique, and as such is governed by a generalized Appleton-Hartree equation~\citep{bittencourt2013fundamentals}. The normal component of the wave index $\mathcal N'$ \eqref{eq:transmited_group_angle_lab} can then be shown to verify the quartic equation
\begin{equation}\label{eq:appleton}
    \Lambda \mathcal N'^4+\Theta \mathcal N'^3 + \Gamma \mathcal N'^2+\Upsilon \mathcal N'+\Xi =0
\end{equation}
with
\begin{subequations}
\label{eq:appleton_coeff}
\begin{align} 
    \Lambda &= P\cos^2(\Psi') + S\sin^2(\Psi'), \\
    \Theta &= {\ntx}'\sin(2\Psi')(P-S), \\
    \Gamma &= \left[\cos(2\Psi')(LR-PS)-LR+2{{\ntx}'}^2(P+S)-3PS\right]/2, \\
    \Upsilon &= {\ntx}'\sin(2\Psi')\left[LR+{{\ntx}'}^2(P-S)-PS\right]/2, \\
    \Xi &= P\left[{{\ntx}'}^2\sin^2(\Psi')({{\ntx}'}^2-S)+LR-{{\ntx}'}^2S\right] + {{\ntx}'}^2\cos^2(\Psi')({{\ntx}'}^2S-LR),
\end{align}
\end{subequations}
where ${\ntx}'={\ktx}'c/\omega'$ and $P$, $L$, $R$, $S$ are the classical functions defined by \cite{stix1992waves}. They here depend on the rest-frame frequency $\omega'$ and explicitly write
\begin{subequations} \label{eq:stix}
\begin{equation}
    P(\omega') = 1-\sum_s \frac{\omega_{ps}^2}{\omega'^2},
\end{equation}
\begin{equation}
     R(\omega') = 1-\sum_s \frac{\omega_{ps}^2}{\omega'(\omega'+\Omega_{cs}')}, \qquad L(\omega') = 1-\sum_s \frac{\omega_{ps}^2}{\omega'(\omega'-\Omega_{cs}')},
\end{equation}
and
\begin{equation}
    S(\omega') = \frac{1}{2}(R+L) = 1-\sum_s \frac{\omega_{ps}^2}{\omega'^2-\Omega_{cs}'^2}.
\end{equation}
\end{subequations}

Compared to standard textbook expressions, the odd terms in \eqref{eq:appleton} are here non-zero because the background magnetic field $\mathbf{B}_0'$ is not aligned with a basis vector in $\Sigma'$. Nonetheless, the quartic equation \eqref{eq:appleton} must similarly yield two modes that are either purely propagative ($\mathcal N' >0$) or evanescent (${\mathcal {N'}^2 <0}$), as usual in a magnetized plasma. Although cumbersome, this quartic equation \eqref{eq:appleton} can be solved to obtain the index $\mathcal N_\pm'$ of these two modes, denoted here by the subscript $\pm$. These wave indexes can then be used to compute the drag experienced by each of the beams using \eqref{eq:transmited_group_angle_lab}. 

\subsubsection{Magnetic field along the direction of motion}

\begin{figure}
    \centering
    \includegraphics[]{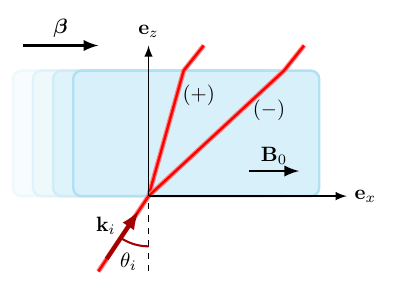}
    \caption{\label{fig:plasma_config_b} The incident wave is in the plane formed by the background magnetic field $\mathbf{B}_0'=\mathbf{B}_0$ permeating the magnetized plasma and the direction of motion $\boldsymbol{\beta}$. The rest-frame modes are the two solutions $(+)$ and $(-)$ for oblique propagation.}
\end{figure}

Rather than going this route, we focus here on the particular case $\Psi'=\pi/2$, that is to say to the case where the magnetic field is parallel to the direction of motion $\vec v$. We note that this configuration, which is illustrated in figure~\ref{fig:plasma_config_b}, could for instance be thought of as a simplified model for the effect of toroidal rotation in a tokamak. In this case~\eqref{eq:appleton_coeff} then gives $\Theta=\Upsilon=0$, so that the Appleton equation \eqref{eq:appleton} reduces to the more usual bi-quadratic equation
\begin{subequations}
\begin{equation}
    \Lambda_\parallel \mathcal N'^4 + \Gamma_\parallel \mathcal N'^2+\Xi_\parallel =0
\end{equation}
where
\begin{align} 
    \Lambda_\parallel &= S, \\
    \Gamma_\parallel &= {{\ntx}'}^2(P+S)-(LR+PS), \\
    \Xi_\parallel &= P({{\ntx}'}^2-L)({{\ntx}'}^2-R).
\end{align}
\end{subequations}
As expected the odd terms are here zero as $\mathbf{B}_0'$ is now along $\mathbf{\hat{x}}$. Also, from the Lorentz
transformations of fields \eqref{eq:field_lorentz_transform}, the lab-frame electric field is null, and the external magnetic field is the same in the lab-frame and in the rest-frame, i.e. $\mathbf B_0 =\mathbf B_0'$.  As a result the prime on the cyclotron frequencies can be dropped in \eqref{eq:stix}.

Although $\bar{n}_{\alpha}'$ still depend on the wavevector of the refracted wave $\vec k_t'$ since $\vec k_t'$ is in general inclined with respect to $\mathbf{B}_0'$, this bi-quadratic equation has, compared to \eqref{eq:appleton}, simpler analytical solutions in the form of
\begin{equation} \label{eq:solu}
    \mathcal N_\pm' = \left[\frac{1}{2\Lambda_\parallel}\left(-\Gamma_\parallel \pm \sqrt{\Gamma_\parallel^2-4\Lambda_\parallel\Xi_\parallel}\right) \right]^{1/2}.
\end{equation}
The two modes denoted here by $(+)$ and $(-)$ are the standard solutions for oblique propagation, also referred to as the slow and fast mode, respectively. In the particular case of normal incidence for which $\vec k_t'\perp\mathbf{B}_0'$, the $(+)$ or slow solution is found to reduce to the O-mode, whereas the $(-)$ or fast solution reduces to the X-mode. These general solutions \eqref{eq:solu} can then be used in \eqref{eq:transmited_group_angle_lab} to derive explicit formulas for the Fresnel drag.

\begin{figure}
    \centering
    \includegraphics[width=1\linewidth]{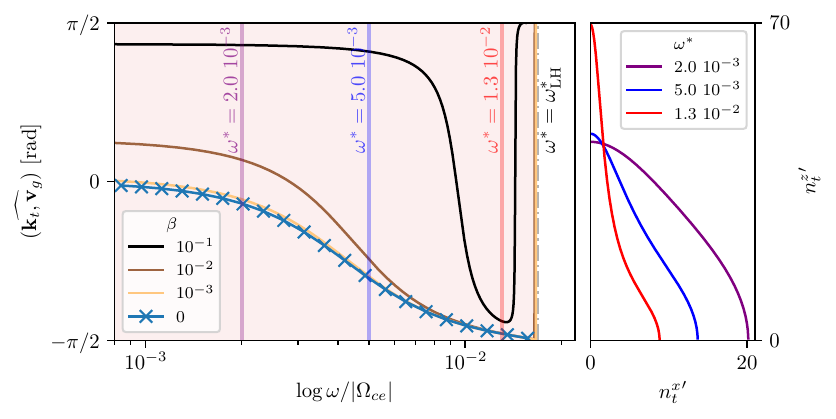}
    \caption{Angle between the transmitted group velocity $\mathbf v_g$ and the wavevector $\mathbf k_t$ as a function of the frequency for a ($-$) mode incident with $\theta_i=-45^\circ$ for different values of the velocity (left), and rest-frame dispersion diagram $({n_t^x}',{n_t^z}')$ for the three wave-frequency highlighted on the left hand side (right). The region of interest here is in between the ion cyclotron frequency ($\Omega_{ci}/|\Omega_{ce}|=5.4~10^{-4}$) and the lower-hybrid frequency ($\omega_{\textrm{LH}}/|\Omega_{ce}|=1.7~10^{-2}$). The plasma parameters are those already used in figure~\ref{fig:Xmode}, leading to $c/v_A=43.4$.}
    \label{fig:Planarmode45}
\end{figure}

To illustrate how drag effects and rest-frame anisotropy can compete with one another, figure~\ref{fig:Planarmode45} plots on the left hand side the same angle $(\widehat{\vec k_t,\vec v_g})$ as in figure~\ref{fig:Xmode}, but we consider now the $(-)$ mode at finite incidence angle $\theta_i=-\pi/4$. We focus here on the low-frequency branch. For frequencies just above the ion-cyclotron frequency, we observe a behavior similar to that of the compressional Alfv{\'e}n branch at normal incidence already observed in figure~\ref{fig:Xmode}, that is a drag in the direction of motion that increases with the velocity, and that can be significant even for modest $\beta$. This similarity can be explained as follows. For $\omega'\sim\Omega_{ci}$ one shows that the axial wave index $\mathcal N_{-}'$ is large for as long as 
\begin{equation}
{n_t^x}'\ll \sqrt{R}\sim\frac{\omega_{pi}}{\sqrt{2}\Omega_{ci}},
\end{equation}
reaching
\begin{equation} 
    \mathcal N_{-}' \sim \sqrt{\frac{LR}{S}}\sim\frac{\omega_{pi}}{\Omega_{ci}}
\end{equation}
for perpendicular propagation ${n_t^x}'=0$. This is verified in the dispersion diagram on the right hand side in figure~\ref{fig:Planarmode45}. Since ${n_t^x}'\leq 1$ from \eqref{eq:snell_field}, this shows that the rest-frame refractive index will be large, which from Snell's law implies that the refracted wavevector $\vec k_t'$ is close to $\mathbf{\hat{z}}$. This in turn implies nearly perpendicular propagation in $\Sigma'$, thus the X-mode like behavior. 

As the frequency increases, however, we observe a departure from this behavior, with the angle $(\widehat{\vec k_t,\vec v_g})$ that now decreases with frequency. It notably becomes negative for large enough frequency, and that even for significant $\beta$. Because the rest-frame refractive index remains large (see the the right hand side in figure~\ref{fig:Planarmode45}), it implies that the component of the group velocity along $\vec v$ must now be negative. The reason for that, as supported by the curve obtained for $\beta=0$ in figure~\ref{fig:Planarmode45}, is the rest-frame anisotropy. Indeed, the angle between $\vec k_t'$ and $\vec v_g'$ grows and approaches $-\pi/2$ at the lower-hybrid resonance.  We see in figure~\ref{fig:Planarmode45} that, short of very large $\beta$, the anisotropy progressively suppresses drag effects in the frequency range below the lower-hybrid frequency. Yet, drag effects eventually dominate again in the immediate vicinity of the cutoff, as the group index goes to infinity at the resonance. This translates into a sudden $\pi$ upshift near $\omega=\omega_{\textrm{LH}}$ in figure~\ref{fig:Planarmode45}. 

For completeness, we note here (not shown in figure~\ref{fig:Planarmode45}) that a behavior similar to that discussed earlier for the X-mode, notably enhanced drag and dispersion effects near high frequency cutoffs and resonances (electron cyclotron range), is also observed in this configuration.

In summary, rest-frame anisotropy is found to bring about additional complexity on top of the motion drag-effects already identified for the X-mode. These two effects can notably oppose one another, with a relative importance that depends strongly on the wave frequency. More practically, the fact that these results are obtained for a configuration which, while very simplified, in essence matches that of a toroidal flow in a tokamak, and for a wave frequency range relevant to magnetic confinement fusion applications, points to the need to explore these manifestations further.

\section{Conclusions} \label{sec:conclu}

The transmission angles for the wavevector and the group velocity of a wave at oblique incidence on an anisotropic medium in uniform linear motion directed along the interface, as observed by an observer in the lab-frame, have been determined analytically. These findings confirm and extend results that had been previously established, notably for isotropic medium and/or for normal incidence.

These lab-frame relations for the wavevector and group velocity, which were obtained by considering Snell's law in the reference frame in which the anisotropic medium is at rest, are further shown to be consistent with generalized Snell's law. More specifically, the interface relations derived here match Snell's law written for the rest-frame index one obtains by invoking the covariance of the dispersion relation between inertial frames. In doing so one can then retire the need for the change of frame of reference in determining the lab-frame relations, by considering instead the moving medium has an equivalent medium bestowed with motion-dependent properties. This direct method may however come at the expense of physical insights.

These new results were then applied to examine the effect of motion on waves incident on a magnetized plasma in uniform linear motion. Starting with the simpler case where the magnetic field is normal to the incident wavevector, for which the rest-frame wavevector and group velocity of the refracted wave are aligned, we show that while the O-mode is unaffected by the motion, the motion can in contrast affect the X-mode. While, as previously noticed, drag effects are weak at high frequency, it is found here that they could be significant for the low frequency compressional Alfv{\'e}n branch. Motion, through the Doppler shift experienced by the wave, is also found to create asymmetric total reflection conditions, and even incidence-angle dependent propagation bands near cutoffs and resonances. 

We finally considered the case where the magnetic field is aligned with the medium's direction of motion, for which the rest-frame wavevector and group velocity of the refracted wave are no longer aligned. While very simplified, this configuration share similarities with the geometry of a wave incident on toroidally rotating tokamak plasma. In this case it is found that, in addition to the effects found for perpendicular propagation, anisotropy and drag can now compete with one another, notably near the lower-hybrid frequency.

Looking ahead, the finding that plasma motion can under certain conditions affect the trajectory of waves in possibly non negligible ways, notably for wave frequencies below the electron cyclotron frequency, confirms that accounting for these effect could be important for the accurate modeling of radiofrequency waves in magnetic confinement fusion plasmas. With that in mind, a goal with these results in hand would be to quantify how large this motion-induced wave trajectory corrections are in practical configurations, similarly to what has been done for instance for the corrections due to spin-orbit coupling~\citep{Fu2023}.

\section*{Acknowledgments}

This work was supported by the French Agence Nationale de la Recherche (ANR), under grant ANR-21-CE30-0002 (project WaRP). It has been carried out within the framework of the EUROfusion Consortium, funded by
the European Union via the Euratom Research and Training Programme. The views and opinions expressed herein do not necessarily reflect those of the European Commission. JL acknowledges the support of ENS Paris-Saclay through its Doctoral Grant Program.


\end{document}